# What the Boltzmann money game teaches us about statistical mechanics (and maybe economics).

Dmitrii E. Makarov*

Department of Chemistry and Institute for Computational Engineering and Sciences,

University of Texas at Austin, Austin, Texas 78712, USA

## Abstract

This note explains why a large class of fair, or reversible "money games", i.e., stochastic models of wealth redistribution among agents, lead to steady states described by canonical and microcanonical distributions. The games considered include, for example, ones where more than two agents can be simultaneously involved in money transfers (similarly to many-body collisions in chemical kinetics) and where amounts transferred between agents are random. At the same time, money games that break time reversal symmetry can also lead to the canonical/microcanonical distributions, as illustrated by an explicit example.

*makarov@cm.utexas.edu

Many physical chemistry instructors use the "money game" to help undergraduates see how the Boltzmann distribution emerges from basic statistical principles[1]. In my Department at the University of Texas at Austin, we learned this pedagogical device from the late John F. Stanton, who was my colleague for many years. As the Boltzmann distribution also often applies to wealth, money games are also prominent models in econophysics[2]. The particular model discussed here is an extension of the game known in econophysics as the Bennati-Dragulescu-Yakovenko model[3, 4], but here the term "Boltzmann money game" used by John is adopted. Our work on money games – fascinating models of equilibrium and nonequilibrium statistical mechanics with relevance to physical chemistry – was motivated by discussions with John, who provided valuable feedback on many occasions. This note explains why such money games inevitably lead to the Boltzmann distribution for a very broad class of game rules, where, for example, the players can occupy different spatial locations preventing them from trading directly with distant partners, or where multiple players may be involved in a single transaction. This is analogous to thermal equilibration of physical systems that are not uniform materials or nearly ideal gases. The only essential rule required is that the game is fair (to be explained bellow). This discussion generalizes earlier proofs of the Boltzmann distribution as a steady-state distribution for more limited classes of games[5, 6].

To a physicist or a chemist, the money game may serve as a metaphor for a large collection of atoms or molecules exchanging energy via collisions. In its most basic form, one envisions $N$ players (particles) sharing $E$ coins (this particular notation is to remind us that coins are analogous to energy quanta here). Upon a random encounter of two players, a coin is transferred between them. The game is fair in that this wealth transfer is not biased toward a richer or poorer player (a more precise definition of fairness will be given later), but a player who is broke can only receive a coin (debt is not allowed). The game can be represented as a random walk on a lattice shown, for just $N = 3$ players and for $E = 3$, in Fig.1. In this figure, each lattice edge corresponds to a transition where one coin is transferred. Each lattice vertex corresponds to a ("microscopic") state $(\epsilon_1, \epsilon_2, \ldots, \epsilon_N)$ where $\epsilon_i$ is the content of the $i$-th player's wallet, with the obvious requirement that those must add up to $E$, just like the energies of the particles adding up to a conserved quantity.

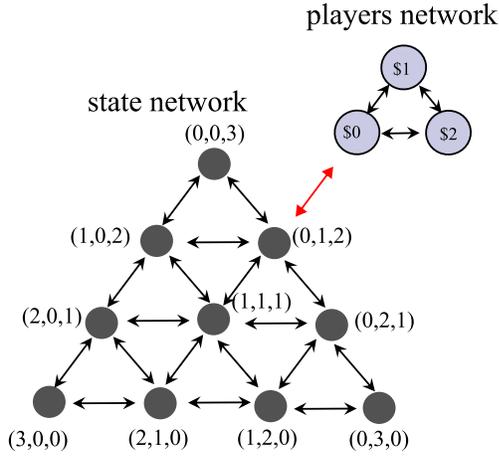

**Figure 1.** Money game with $N = 3$ players sharing a wealth of $E = 3$ coins. The system has 10 possible states, and transitions within the network correspond to transfer of a single coin between pairs of players. The money game is a random walk on a network (graph) containing the 10 states. This network is not to be confused with the network of connections between the players: in the case depicted here every pair of players is connected (i.e., each player can exchange coins with each other player). The rules of the game are specified by the transition probabilities between the graph vertices: some of those may be zero, and self-transitions (not shown here) may also be possible, where the state of the system remains unchanged.

In statistical mechanics, we are usually interested in the thermodynamic limit, $N \gg 1$. Because this limit is difficult to represent pictorially, the parameters in Fig. 1 do not satisfy this condition.

A central quantity of interest for us is the distribution of wealth, i.e., the probability $P_i(\epsilon_i)$ that the $i$-th player has $\epsilon_i$ coins. This is the marginal probability

$P_i(\epsilon_i) = \sum_{j \neq i} p(\epsilon_1, \ldots, \epsilon_i, \ldots, \epsilon_N),$ (1)

where $p(\epsilon_1, \ldots, \epsilon_i, \ldots, \epsilon_N)$ is the joint probability distribution for a "microstate" of the game specified by the amounts owned by each player. If all players are equivalent, this probability is the same for every player, and so the index $i$ can be dropped, $P_i(\epsilon_i) \equiv P(\epsilon_i)$. Just like in molecular systems obeying the Boltzmann distribution, in many versions of the many game $P(\epsilon_i)$ turns out to be an exponential function of wealth $\epsilon_i$. In the broader context of statistical mechanics, textbook proofs[7-9] of this result usually involve two steps.

**Step 1.** One way or another, Laplace's principle of maximum indifference[10] is invoked, which implies that all allowable "microscopic" states $(\epsilon_1, \epsilon_2, \ldots, \epsilon_N)$ are equally probable.

That is,

$$p(\epsilon_1, \epsilon_2, \ldots, \epsilon_N) = \begin{cases} \frac{1}{\Omega(E,N)}, & \sum_i \epsilon_i = E \\ 0, & \sum_i \epsilon_i \neq E \end{cases}, \qquad (2)$$

where $\Omega(E, N)$ is the total number of microstates for $N$ players satisfying the constraint of their energies adding up to $E$. For the game shown in Fig.1, $\Omega(E, N)$ is the number of vertices of the lattice representing the game. Eq. 2 is plausible but usually not provable from first principles (e.g., Newtonian dynamics laws) except in special cases or under additional assumptions[8] – it is therefore usually taken as a postulate of statistical mechanics. Eq. 2 maximizes the entropy of the microscopic system and is known as the microcanonical distribution.

**Step 2.** Suppose a player has $\epsilon$ coins. Then the remaining $N - 1$ players share $E - \epsilon$ coins. There are $\Omega(E - \epsilon, N - 1)$ ways to redistribute the money among them. Assuming Eq. 2, it is clear that the probability $P(\epsilon)$ is given by

$$P(\epsilon) = \frac{\Omega(E-\epsilon, N-1)}{\Omega(E,N)} \qquad (3)$$

Eq. 3 counts the fraction of all possible microscopic states under the constraint that one player has a fixed number of coins. To calculate $\Omega(E, N)$ we use a known trick: lay out $E$ coins along a line and partition them with $N - 1$ sticks, such that the money amount between two neighboring sticks represents a player's wealth. We thus have $E + N - 1$ objects on a line, of which $E$ objects are coins and $N - 1$ objects are sticks. The number of such arrangements is equal to the number of ways $E + N - 1$ items can be placed into two baskets, one designated for the sticks and the other for the coins, which is equal to the binomial coefficient

$$\Omega(E, N) = \frac{(E+N-1)!}{E!(N-1)!} \qquad (4)$$

We will come back to our trick with sticks partitioning the coins at the end of this paper. For now, we consider the case of a very large number of players and large pool of money, $N \gg 1, E \gg 1$, and fix the average wealth per player at $\langle \epsilon \rangle = E/N$. Using Stirling's approximation for the factorials in Eq. 4, (e.g., $\ln N! \approx N \ln N - N$), and assuming $\epsilon \ll N, E$, one finds from Eqs. 3 and 4:

$$P(\epsilon) \approx \frac{1}{1+\langle \epsilon \rangle} \left(\frac{\langle \epsilon \rangle}{1+\langle \epsilon \rangle}\right)^\epsilon = \frac{1}{1+\langle \epsilon \rangle} e^{-\epsilon/T}, \qquad (5)$$

which is the Boltzmann distribution with a temperature of

$$T = \frac{1}{\ln(1+\frac{1}{\langle \epsilon \rangle})} \quad (6)$$

Step 2 assures us that the Boltzmann distribution follows from equiprobability of microscopic states, Eq. 2, assuming a large number of players and ignoring the probabilities of very rare events where each player accumulates large fractions of the total money pool (that is, Eq. 5 assumes $\epsilon \ll E$). For example, Eq. 5 is obviously violated for $\epsilon > E$, but both the true probability of such an event (equal identically to zero) and the prediction of Eq. 6 are negligible in this case anyway.

We now focus on Step 1. In contrast to the case of systems consisting of interacting atoms, the equiprobability of all the states, Eq. 2, need not be postulated for money games, as ***it can be derived from the rules of the game***. This has been done for several versions of the money game, with the most general case considered in refs[5, 6]. In what follows I will prove this for an even more general class of money games, thus highlighting the fundamental reason behind the validity of Eq. 2, and therefore of the Boltzmann distribution of wealth. Consider an elementary step in a money game, in which two players, numbered $i$ and $j$, with their respective funds $\epsilon_i$ and $\epsilon_j$, engage in a transaction. Let $p(\ldots \epsilon'_i \ldots \epsilon'_j \ldots | \ldots \epsilon_i \ldots \epsilon_j \ldots)$ be the conditional probability that the wealth of the first player becomes $\epsilon'_i$ while the wealth of the second becomes $\epsilon'_j$. To simplify the notation, enumerate all possible states $(\epsilon_1, \ldots, \epsilon_N)$ of the game's network by a single index $\alpha$ (e.g., $\alpha$ ranges between 1 and 10 for the $N = E = 3$ game in Fig. 1) and call $p(\alpha)$ the probability of the system to be in some particular state $\alpha$ attained after playing the game for a long time (i.e., the steady-state probability). Further denote $p(\alpha'|\alpha)$ the conditional probability that a transaction starting in $\alpha$ will lead to a new state $\alpha'$. We will assume that the network formed by the states is irreducible, which means that any two of its states are connected by some path; this rules out scenarios where isolated players or groups of players cannot exchange coins with other groups. Then there is a unique steady state of the game, where all $p(\alpha)$'s remain constant. In this state, these probabilities satisfy the equation

$$p(\alpha) = \sum_{\alpha'} p(\alpha|\alpha')p(\alpha') \quad (7)$$

This equation literally means that the system in state $\alpha$ has arrived from some state $\alpha'$ in one step of the game; for a given $\alpha'$ the probability of this is $p(\alpha|\alpha')p(\alpha')$, which is summed over all previous states $\alpha'$ to get the probability of ending up in $\alpha$. If all the transition probabilities $p(\alpha'|\alpha)$ are known, solving Eq. 7 becomes a linear algebra problem: Let **p** be a column vector with components $p(\alpha)$ and let **T** be a "transition" matrix whose elements are $T_{\alpha'\alpha} \equiv p(\alpha'|\alpha)$. Then Eq. 7 states

**p** = **Tp**, (8)

which means that **p** is a right eigenvector of **T** with an eigenvalue of 1. The transition matrix **T** has an important property that all of its columns add up to 1. Indeed, this is simply the statement that the probability that the game starting in $\alpha$ ends up in any state $\alpha'$ is 1:

$$\sum_{\alpha'} p(\alpha'|\alpha) = 1. \quad (9)$$

This is equivalent to saying that **T** has a left eigenvector given by $\mathbf{v} = (1, \ldots, 1)$ with its eigenvalue equal to 1:

$$\mathbf{vT} = \mathbf{v} \quad (10)$$

Now, if **T** is a symmetric matrix

$$p(\alpha'|\alpha) = p(\ldots \epsilon_i' \ldots \epsilon_j' \ldots | \ldots \epsilon_i \ldots \epsilon_j \ldots) = p(\ldots \epsilon_i \ldots \epsilon_j \ldots | \ldots \epsilon_i' \ldots \epsilon_j' \ldots) = p(\alpha|\alpha'), \quad (11)$$

then the transposed vector $\mathbf{v}^T$ is also its right eigenvector, $\mathbf{Tv}^T = \mathbf{v}^T$. Thus the column vector $\mathbf{p} = \Omega^{-1}\mathbf{v}^T$ with all of its components equal is a right eigenvector satisfying Eq. 8. Here the factor $\Omega^{-1}$ was introduced to ensure that all the probabilities add up to 1. That is, Eq. 11 implies that the steady-state probabilities of all possible microscopic states of the game are equal, which is Eq. 2! From an economics standpoint, Eq. 11 is a precise statement about the fairness of the game – coins are merely shuffled between the players, with each transaction having no preferred direction. From a physics perspective, on the other hand, Eq. 11 is reminiscent of the statement of microscopic reversibility in a Newtonian system: for every pair of atoms exchanging energy in a collision there is a reversed trajectory with the same energy transferred backwards, which is obtained through a reversal of atom's velocities; but of course, a direct comparison of Newtonian dynamics with stochastic dynamics of money games is not possible.

The symmetry of the transition matrix **T** expressed by Eq.11 is the essential feature of the various versions of the money game that lead to the Boltzmann statistics[2, 4-6, 11-14], Eq. 5. In the theory of Markov chains, matrices like this are known as "doubly stochastic", which means that both their columns and rows add up to 1. Importantly, this symmetry is satisfied by a broad class of games. Imagine, for example, an arbitrary network of interconnected players, such as the one in Fig. 2. The money can be transferred along any edge of the network, and the only condition we require is that any transfer is reversible: the probability of moving $\Delta\epsilon$ coins along any edge connecting two players is the same in each transfer direction. A simple example of such a reversible game is when each transaction between connected players randomizes their wallet contents while preserving the total. That is, we have $\epsilon_i' + \epsilon_j' = \epsilon_i + \epsilon_j$, with $\epsilon_i'$ and $\epsilon_j'$ chosen randomly at each exchange; the outcome of the exchange depends on the combined wealth of the players pair but not on their individual wealth.

As another example, suppose that at every step a pair of connected players on the network is picked at random. The selected players, $i$ and $j$ use a random number generator to decide on the number of coins to be transferred, $\Delta = \epsilon_i - \epsilon_i' = \epsilon_j' - \epsilon_j$, from a symmetric distribution $\rho(\Delta) = \rho(-\Delta)$. Now, suppose the player who is losing money in this trade (say the $i$-th player) requests the transfer from their bank, but the bank discovers that there are no sufficient funds in the account, i.e. $\epsilon_i - \Delta < 0$, which is not allowed by the game. Then the transaction is rejected and the state $\epsilon_i' = \epsilon_i, \epsilon_j' = \epsilon_j$ is counted as the new state. In other words, there are self-transitions in the network, in which the state does not change (see Fig. 1). Those are described by the diagonal elements of the matrix $p(\alpha|\alpha)$, and they do not affect the symmetry of the matrix. It is easy to see that the conditional probability for each move with nonzero money transferred is given by $p(\alpha'|\alpha) = \rho(\Delta) = \rho(-\Delta) = p(\alpha'|\alpha)$, satisfying Eq. 11.

We conclude that the Boltzmann distribution will be achieved, for such a fair game, for an arbitrary connected network. For example, we could compare the cases of "global trade" with every player trading with every other player and local trade, where players form clusters (e.g. cities or countries) of frequent transactions with infrequent exchanges between clusters. While the dynamics of such models could be quite different, the equilibrium distribution will be the same. Indeed, the Boltzmann distribution was found in simulations and theoretical studies of trading within networks of varied topologies[4-6, 11-14]. In the language of statistical mechanics, no matter what the network topology is, the economics of its different parts will equilibrate to the same "temperature".

Two comments are due here. First, the "fairness" condition, Eq. 11, is stronger than the condition of temporal reversibility of the sequence of states visited by the network (i.e. time-reversibility of the Markov chain). The former is satisfied if the network satisfies detailed balance for every pair of states,

$p(\alpha'|\alpha)p(\alpha) = p(\alpha|\alpha')p(\alpha')$, (12)

which can be valid even if $p(\alpha)$'s are not the same. An example is provided by a recent model, in which some of players are cheaters claiming, with some probability, that they have zero wealth thus preventing their loss of money in an exchange[15, 16]. In a transaction between a cheater and a non-cheater, the odds are in favor of the cheater, making Eq. 11 invalid. As a result, the overall wealth distribution is no longer exponential. The detailed balance condition is nevertheless satisfied by this game[15]. Second, the above discussion shows that special care should be taken when specifying the rules of the game in cases when one of the traders does not have enough money for the exchange, particularly when they have zero wealth. In the above example where the transferred amount is random, validity of Eq. 11 is ensured by introducing steps in which the transfer is rejected and the state does not change. In ref.[17] we have considered a slightly different rule, in which a money transfer takes place with certainty in the allowed direction. As a result, Eq. 11 was violated for some states (those with at least one of the trading partners having zero money), resulting in a deviation from Eq. 5, particularly for the case $\epsilon = 0$. When wealth is abundant, $\langle \epsilon \rangle = \frac{E}{N} \gg 1$, we have, according to Eq. 5, $P(0) \approx \frac{N}{E+N} \approx \frac{1}{\langle \epsilon \rangle + 1} \ll 1$. Then encounters where one of the participants is broke are rare, and Eq.5 is still satisfied approximately, as observed in ref.[17].

In addition to generalizing earlier money games to allow transfer of arbitrary amounts of money, we can also allow transactions involving any number of players. Indeed, there is nothing in Eq. 11 that requires that $\alpha$ and $\alpha'$ differ by only the wallet contents of two players. It is possible that every game event results in wealth redistribution of any number of players, or even all the players in the game. As long as Eq. 11 holds, the resulting distribution will be Boltzmann! In the physics language, many-body interactions do not change the equilibrium distribution.

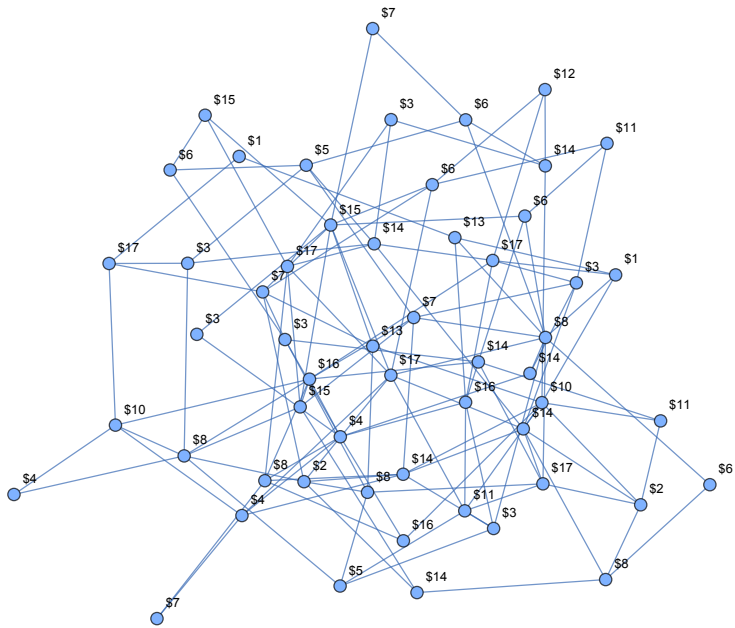

**Figure 2.** A large network of connected players. The number at each vertex indicates a players' wealth – the set of all these numbers determines the "microscopic" state of the game. The money game consists of transfers of coins between pairs of connected players, that is, shuffling of the coins along the edges of the network[5]. If each shuffling move is reversible, the distribution of money on each network node is given exactly by Eq. 3, or, asymptotically by Eq. 5.

While the symmetry of Eq. 11 ensures the Boltzmann distribution for a rather general class of games with arbitrary connectivity among the players, all bets are off if this symmetry is violated[18, 19]. In particular, games that violate detailed balance (Eq. 12) result in very different wealth distributions. In the language of nonequilibrium statistical mechanics, those games have nonequilibrium steady states that break time-reversal symmetry: the sequence of states visited by the game is statistically different from the time-reversed sequence. An example is provided by unfair games rigged in favor of wealthier players, as the time reversals of each coin transfer in such games would appear to favor poorer players. Such games may result in Pareto-type wealth power-law distributions similar to those observed in real-world economies[17]. In contrast to the robust distributions of wealth in fair games, there is no reason to expect that a non-Boltzmann steady-state distribution would be equally robust and independent of the game details such as the connectivity between players.

It is interesting to note, however, that the validity of Eq. 11 is only a sufficient but not a necessary condition for the Boltzmann statistics. An example of a time-reversal-symmetry-breaking game that still leads to Eqs. 2-3, and thus to Eq. 5 in the limit of a large number of players, is shown in Fig. 3(a). In this game all the players are connected into a ring, and random amounts of money are only passed along the ring in the counterclockwise direction. The counterclockwise flow of money breaks time-reversal symmetry (Eqs. 11 and 12), but the game can still preserve the Boltzmann distribution. To see this it is expedient to recall the trick with partitioning coins with sticks: We arrange all the coins in a ring and partition them such that the coins between two adjacent sticks represent the wealth of a player, Fig. 3(a). Each money transfer between two neighbors is equivalent to moving a stick dividing their coins; thus, instead of money flowing counterclockwise we can think of sticks moving randomly in the clockwise direction. If the sticks' dynamics preserves their statistical independence then Eqs. 2 and 3 will still be valid. A slight problem with this scenario is that sticks cannot overtake each other, as that would correspond to the wealth of a player becoming negative, and so there is effective "repulsion" between sticks. A simple way out is to change the rules of the game such that, whenever two sticks pass each other, the state of the system is changed such that the wealth of each player is equal to the number of coins between the sticks taken with the positive sign. In other words, the wealth of each player is *defined* to be the distance between a pair of sticks that move freely along the circle. It can be shown that a simple modification of the reversible game above, in which a player is selected at random to give money $\Delta \geq 0$ to the next player in the counterclockwise direction, with $\Delta$ drawn from some specified distribution and with the transfer rejected whenever $\Delta$ exceeds the wealth of the giver, also results in Boltzmann statistics. In particular, in the numerical example shown in Fig. 3(b), the money transferred (when allowed) is a random integer between 0 and 3.

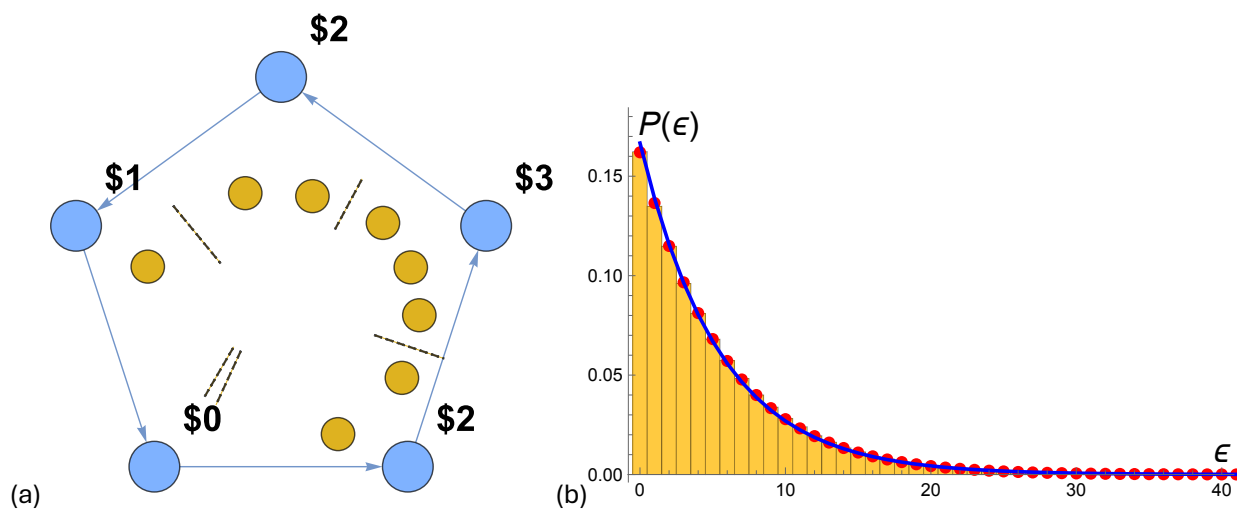

**Figure 3**. (a) Nonequilibrium money game in which money is passed between players in counterclockwise direction. Equivalently, one may think of the players' coins laid out along a ring and partitioned by sticks, so each money transfer results in the stick dividing the coins of two players moving clockwise. (b) Histogram of the probability distribution of wealth from a simulation of such a nonequilibrium money game with $N = 30$ and $E = 150$. Red markers and blue line show analytical results of Eq. 3 and Eq. 5, respectively.

In summary, simple money games in which random encounters between players result in redistribution of wealth offer an example where the fundamental laws of statistical mechanics arise from dynamic laws without having to invoke additional postulates such as equiprobability of microscopic states. They further illustrate that, while reversibility may be a sufficient condition for the Boltzmann distribution, it is not a necessary one. My interest in this problem was ignited by one of the many random encounters with John F. Stanton in a hallway of the Welch Hall at the University of Texas, a testament to the profound impact of John's insight on the work of his colleagues.

I am also grateful to Eric Anslyn, Kristian Blom, Aljaž Godec, Hagen Hofmann, and Kevin Song for comments and discussions and to the National Science Foundation for the financial support via grant CHE 2400424.